\documentclass[onecolumn]{emulateapj}

\shorttitle{On the radial acceleration relation for galaxies}
\shortauthors{C. E. Navia}

\usepackage{amsmath}

\begin{document}


\title
{
On the radial acceleration relation for galaxies in a wide range of redshifts
}


\author{Carlos. E. Navia }
\affil{Instituto de F\'{i}sica, Universidade Federal Fluminense, 24210-346, Niter\'{o}i, RJ, Brazil}


\altaffiltext{1}{E-mail address:navia@if.uff.br}


\begin{abstract}
The thermal history of the Universe is introduced within the Debye Gravitational Theory (DGT), a thermodynamic theory of induced gravity, and allows to obtain the evolution of systems with the redshift. DGT reproduce the ESO VLT observations, showing falling outer rotation curves for galaxies at redshift above 0.77. 
A scaling law is observed in the radial acceleration relation (RAR) of galaxies. For accelerations smaller than $\sim 10^{-10}ms^{-1}$
the observed acceleration $g_{obs}$ decreases more slowly than acceleration generated by the baryonic mass $g_{bar}$, following always the relation $g_{obs} \sim \sqrt{g_{bar}}$.
The RAR does not care about the specific
properties of the galaxy, the relation exists in nearby high-mass elliptical and low-mass spheroidal galaxies.
In this paper, through a straightforward analysis, we show that according to DGT, the RAR scaling law observed in nearby galaxies is broken, when are considered distant galaxies. The extreme case happen for galaxies at redshift above 0.77, because according to DGT, the galaxies have declining rotation curves, and the rotation velocity falling faster than the Keplerian-law curve. Then in this case $g_{obs}$ decreases more faster than $g_{bar}$, this means that $g_{obs}$ is systematically lower than $g_{bar}$.
We show that DGT prediction for the RAR of galaxies at high redshift is in agreement from those obtained from the falling rotation curves observed by VLT telescope.
\end{abstract}




\section{Introduction}
\label{sec:intro}

On 2017, DES infrared telescope, has revealed new measurement of dark matter structure in the universe
\footnote{http://news.fnal.gov/2017/08/dark-energy-survey-reveals-accurate-measurement-dark-matter-structure-universe/}. The DES results show a slight disagreement with the satellite Planck's measurements based on the CMB. Planck reveals around 7\% higher, the fraction of dark matter than DES.

These results contrast with the most successful detection technology implemented in underground detectors to search for the most widely accepted hypothesis on dark matter particles, the WIMP \citep{behn11}.
So far, there is no direct evidence of WIMP or other dark particles.
The three major underground detectors, such as the LUX (USA), Panda (China) and Xenon 1000 (Italy),
have reported negative results \citep{aker16,cui17,apri17}. In addition, so far, there is no evidence of a new physics beyond the standard model, in the Large Hadron Collider (LHC) data \citep{gibn16}.

Thus, the nature of dark matter is still unknown even after the DES results. Apparently, we are under a dichotomy, ``there is a lot of dark matter in the sky, but none in Earth''.
However, the panorama of a sky with a lot of dark matter begins to change, with new measurements, such as the results from dwarf galaxies orbiting around Centaurus A \citep{mull18},  they are rotating in a coherent plane, in the same way, that dwarf galaxies orbit the Milky-Way and the Andromeda galaxy. In disagreement with standard models of galaxy formation. This a clear challenge to cold dark matter cosmology.

 Nearby galaxies are well described by MOdified Newtonian Dynamics (MOND) theory \citep{milg83a,milg83b}. MOND successfully explaining the asymptotic flatness of rotation curves, the baryonic Tully-Fisher relation (BTFR)\citep{tull77}, that correlated the galaxy masses with their radial velocities \citep{mcga11,fama12,sand90}.  Recently, from an accurate Rotation Curve (SPARC) galaxy sample found a scaling-law relation between the radial acceleration inferred from the rotation curves and the acceleration due to the baryonic component \citep{mcga16}. This radial acceleration relation (RAR) is the same relation followed by all galaxies, independent of type, size, and morphology. It has been suggested that this relation may be evidence for new physics, beyondCDM.
 
 It is claimed that the RAR scaling-law will be a consequence of the BTFR \citep{whee18}. Or vice versa, the BTFR can be a consequence of the RAR scaling-law. This would explain why MOND is consistent with the RAR scaling-law. 
However, it is also claimed that  RAR arises as a consequence of the interaction between baryons and dark matter. If galaxies
with a variety of baryonic and dark matter halo profiles, have rotation curves satisfying the BTFR, they satisfy also, the RAR 
scaling-law.
  However, the slope ($\sim 4$) of the BTFR is hard to be obtained within the
framework of dark matter models such as LCDM galaxy-formation simulations, that predict a different
slope ($\sim 3$). It is necessary to introduce the so-called stellar feedback processes,
to reproduce the BTFR \citep{broo12,hopk14}


 In 2017, new observations using ESO's Very Large Telescope (VLT) 
indicate that massive, star-forming galaxies during the peak epoch of galaxy formation, i.e., galaxies at redshift above 0.5, were dominated by baryonic or ``normal'' matter \citep{genz17,lang17}.
The result constrains dark matter models, that claim that the effect of the dark matter is greater in galaxies.
 The result also constrains the so-called modified gravity models or theories such as MOND that mimic the presence of dark matter.

Also on 2017 have been shown that the ESO-VLT results about galaxies at redshifts above 0.5 can be reproduced by the
Debye Gravitational Theory (DGT) \citep{navi17}.  DGT is a thermodynamic gravitational theory, it incorporates the structure of Debye theory of specific heat of solids at low temperatures, within the Entropic Gravitational Theory \citep{verl11}. DGT also incorporates the thermal history of the universe, as observed in the CMB. This framework is useful because
allow obtaining dynamic equations as a function of the redshift. DGT has only a free parameter, the Debye Temperature $T_D=6.35$ K, and that is determined from observations and can be correlated with the acceleration scale parameter.
Indeed, the so-called deep-MONDian regime, appears as a limit of DGT, at very low temperatures
$T<<T_D$ (redshifts close to zero). At high temperatures $T>>T_D$, DGT coincides with the Newtonian regime and in this limit DGT is also compatible with General Relativity Theory \citep{verl11}. DGT is also different from Verlinde emergent gravity \citep{verl17},
where the interaction between baryons and the dark energy mimic the presence of dark matter.

In this paper, we present a straightforward analysis based on the DGT of the RAR of galaxies in a wide range of redshifts. DGT reproduce
the RAR scaling-law observed in nearby galaxies. But to distant galaxies, DGT predicts a breaking of the RAR scaling-law and DGT results are in agreement with observations derived from VLT measurements at high redshift.

This paper is organized as follow.
In section~\ref{DGT} the DGT formalism is briefly reviewed. Section~\ref{rotationa} discusses how well DEGT works in explaining the shape of galaxy rotation in a wide range of redshift. This section is essential to understand DGT prediction on the RAR and that is presented in section~\ref{radiala}, and finally, in section~\ref{conclusions} is presented our conclusions.

\section{Background aspects of DGT}
\label{DGT}

Debye model \citep{deby12} of the specific heat of solids at low temperatures is incorporated in Entropic Gravitational Theory 
\citep{verl11}, the result is called as Debye Gravitational Theory (DGT) \citep{navi17}. The main difference with the 
Entropic Gravitational Theory is the behaviour of the information (bits) stored on the holographic screen, in the first case they do not oscillate, while DGT consider the bits as oscillating quasi-particles on the holographic screen, with a continuous range of frequencies, that cuts off at a maximum (Debye) frequency (temperature), $\omega_D= (k_B/ \hbar) T_D$.

A central argument of the holographic principle considers that each bit of information on the holographic screen carries a $1/2k_BT$ of energy, where T is the screen temperature and $k_B$ is the Boltzmann constant. The number of bits, N, on the screen surface is proportional to the area A  of the screen and expressed as $N=(c^3/G \hbar) A$ \citep{hoof93}, where G is Newton gravitational constant. Taking into account the Principle of  Equipartition  of energy, the specific potential energy on the holographic screen can be written as

\begin{equation}
U=\frac{1}{2}N k_B T .
\label{eq:energy0}
\end{equation}
This equation is the root of Entropic Gravity Theory \citep{verl11}.

However, in order to take into account the oscillation of the quasi-particles (information bits) on the holographic screen and
following the analogy with Debye theory, In DGT is substituted 
$k_BT$  by $k_BT\;\mathcal{D}_1(T_D/T)$, to take into account all range of temperatures, especially the low temperatures

\begin{equation}
U=\frac{1}{2}N k_B T \mathcal{D}_1\left(\frac{T_D}{T}\right).
\label{eq:energy}
\end{equation}
The main difference with the Debye theory is in that the third Debye function, $\mathcal{D}_3(x)$ was replaced by the first Debye function, $\mathcal{D}_1(x)$, because the information bits located on the screen have only a vibrational state along of the gradient of Newton potential and it is assumes that the vibrations follows a continuous range of frequencies. $\mathcal{D}_1$ is defined as
\begin{equation}
\mathcal{D}_1\left(\frac{T_D}{T}\right)= \left(\frac{T}{T_D}\right) \int_0^{T_D/T} \frac{x}{e^x-1}dx,
\label{eq:debye}
\end{equation}
The shape of Debye function reflects the Bose-Einstein statistic formula, used in its derivation. 
  
  IF M represents all mass enclosed by the screen surface, the specific potential energy can be written as $U=Mc^2$.
  The gravitational interaction between M and a particle of mass m emerges when the particle is at a distance $\Delta X$
(close to the Compton wavelength) on the screen. The result is the entropy variation, $\Delta S$ of the screen.
 The so called Bekestain entropy variation \citep{beke73} can be expressed as
\begin{equation}
\Delta S=2\pi k_B \frac{mc}{\hbar} \Delta X.
\label{eq:entropy}
\end{equation} 
these relations, allows to obtain the entropic force defined as $F=T \frac{\Delta S}{\Delta x}$.
The more simple case is for a spherical screen of radius R, ($A=4\pi \;R^2$), in this case the number of bits on the screen
is  $N=(c^3/G \hbar) 4\pi \;R^2$. Combining the above equations, the acceleration of the mass m is

\begin{equation}
a\mathcal{D}_1\left(\frac{T_D}{T}\right)= \frac{GM}{R^2},
\label{eq:main3}
\end{equation}

that has two limits

\[ \mathcal{D}_1\left(\frac{T_D}{T}\right) =
  \begin{cases}
    1       & \quad \text{if } T_D/T << 1 \text{ (high T)}\\
    \pi^2(T/T_D)/6  & \quad \text{if } T_D/T >> 1 \text{ (low T)}.\\
  \end{cases}
\]

The first case (high T) reproduce the second Newton's law, whereas the second limit (low T) can see used to obtain the only free parameter of the theory, the Debye temperature, $T_D$. As in the case of the study of the capacity specific to the solids a low temperature, the Debye temperature can also be obtained from observations.

According to Unruh effect \citep{unru76} there a linear correlation between the temperature and acceleration. Then, the Debye function at low temperatures can be written as
\begin{equation}
\mathcal{D}_1\left(\frac{T_D}{T}\right)=\frac{\pi^2}{6} \left(\frac{T}{T_D}\right)=\left(\frac{a}{a_0}\right),
\label{eq:DebyeLow}
\end{equation}

\begin{figure}
\vspace*{-0.0cm}
\hspace*{0.0cm}
\centering
\includegraphics[width=11.0cm]{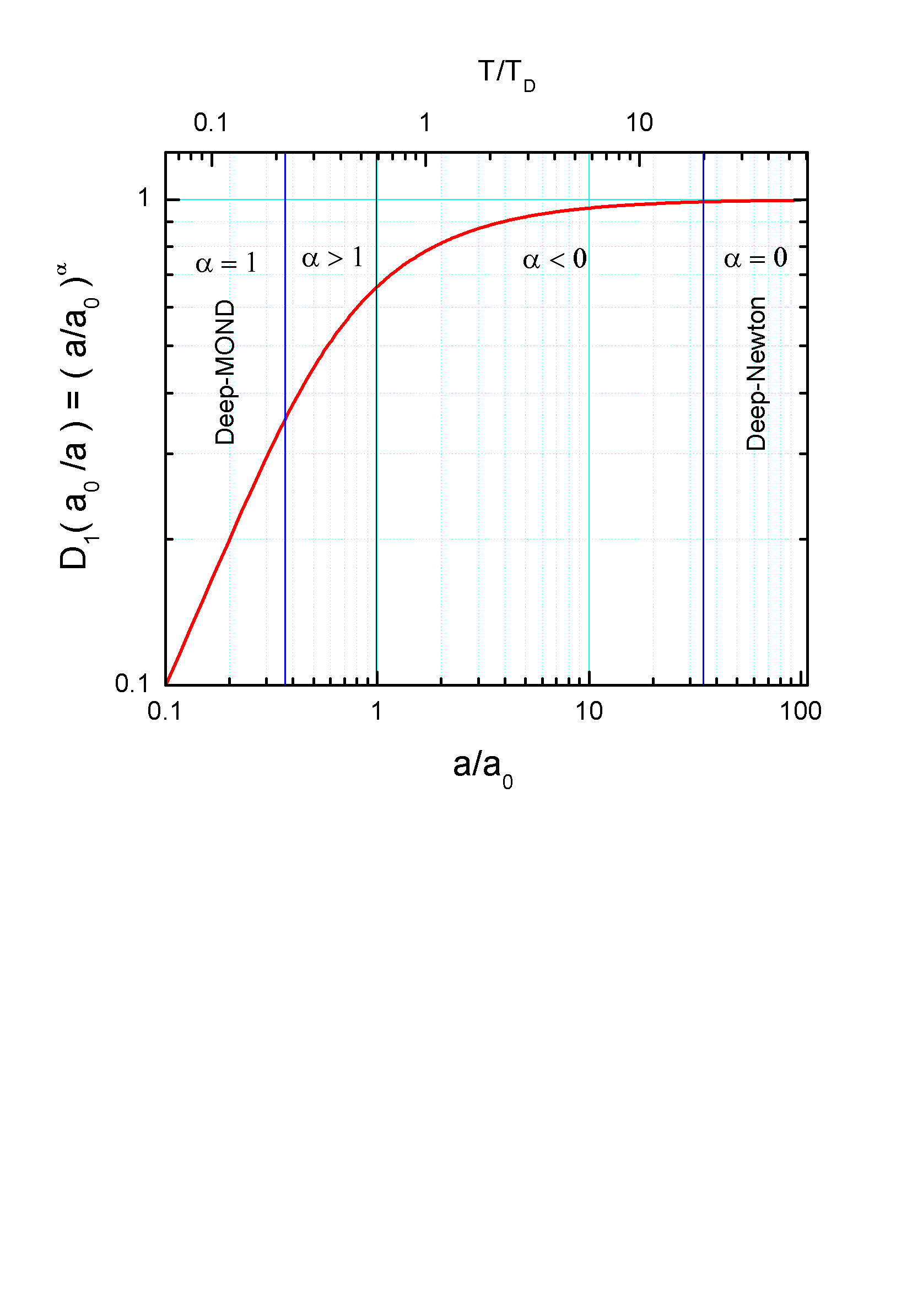}
\vspace*{-6.0cm}
\caption{Debye first function, as a function of the acceleration ratio (bottom scale) and the temperature ratio (top scale), parametrized as a power law.
}
\vspace*{0.5cm}
\label{debye_function}
\end{figure}

we are calling $\pi^2/(6\;T_D)$ as $1/a_0$. Thus, there is a bond between the Debye temperature, $T_D$, and the acceleration scale parameter, $a_0$, and the limit at low temperatures
of the Debye function can be written as

\begin{equation}
\mathcal{D}_1\left(\frac{T_D}{T}\right)=\frac{a}{a_0},
\label{eq:limit}
\end{equation}
and the Eq.~\ref{eq:main3} can be written as
\begin{equation}
a(\frac{a}{a_0})=\frac{GM}{R^2}.
\label{eq:main2}
\end{equation}
This  equation is the root of the MOND theory \citep{milg83a}.

The distributions of a wide variety of physical phenomena, approximately follow a power law over a wide range of magnitudes
A power law is scale invariant, i.e., it is independent of the initial size of the two quantities that correlated. Thus, for all range the temperatures, the Debye function can be parametrized by a power function of 
type $(a/a_0)^{\alpha}$,  and  Eq.~\ref{eq:main2} becomes
\begin{equation}
a\left(\frac{a}{a_0}\right)^{\alpha}= \frac{GM}{R^2}.
\label{eq:mainDGT}
\end{equation}
This equation is the root of DGT. Fig.~\ref{debye_function} shows the first Debye function parametrized as a power-law as a function of acceleration ratio (bottom scale) and temperature ratio (top scale). The two asymptotic  cases of Eq.~\ref{eq:mainDGT} are: 

\[ a\left(\frac{a}{a_0}\right)^{\alpha} =
  \begin{cases}
    a       & \quad \text{if } \alpha = 0 \text{ (high T- deep-Newtonian )}\\
    a\left(\frac{a}{a_0}\right)  & \quad \text{if } \alpha = 1 \text{ ( low T - deep-MONDian)}.\\
  \end{cases}
\]

These two limits show that $\alpha$ also depends on temperature. In general, the $\alpha$ dependence with the temperature is obtained as following. The Eq~\ref{eq:mainDGT} means that
 \begin{equation}
 \mathcal{D}_1\left(\frac{a_0}{a}\right)=\left(\frac{a}{a_0}\right)^{\alpha},
 \end{equation}
 that allow to obtain a expression to the index $\alpha$
 \begin{equation}
 \alpha = \frac{\log \mathcal{D}_1\left(\frac{a_0}{a}\right)}{\log \frac{a}{a_0}}=
          \frac{\log \mathcal{D}_1\left(\frac{6\;T_D}{\pi^2T}\right)}{\log \frac{\pi^2T}{6\;T_D}},
 \label{alpha_a}
 \end{equation}
 
\begin{figure}
\vspace*{-0.0cm}
\hspace*{0.0cm}
\centering
\includegraphics[width=16.0cm]{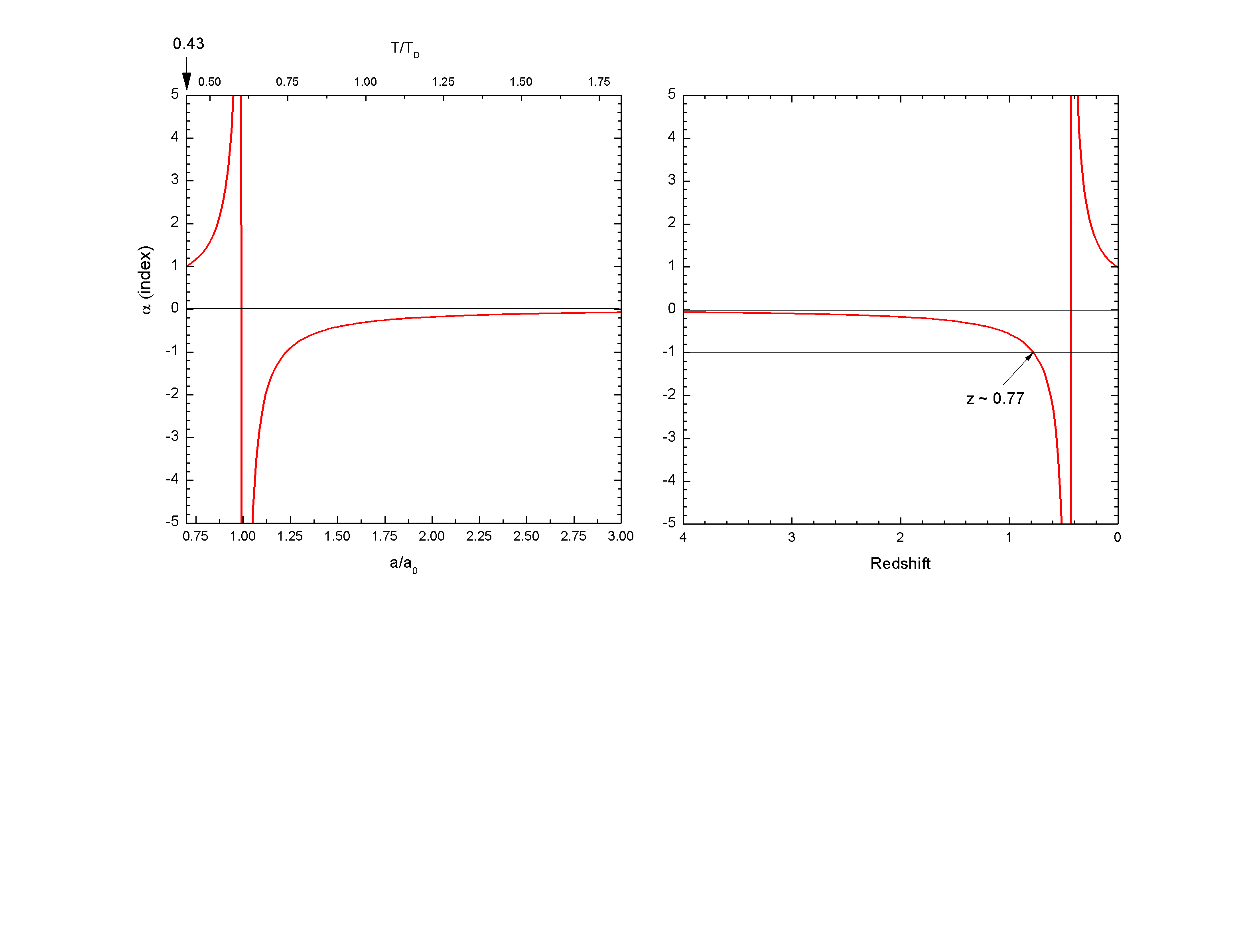}
\vspace*{-5.0cm}
\caption{Left: The $\alpha$ index as a function of acceleration ratio(bottom scale) and the temperature ratio (top scale). The vertical arrow at top left corner,represent the temperature ratio $T /T_D$ at $\alpha=1$, and correspond to z=0. Right: The $\alpha$ index  as a function of the redshift.
}
\label{alpha_redshift}
\end{figure}

Fig~\ref{alpha_redshift}(left panel) shows the dependence of $\alpha$ as a function of acceleration ratio $a/a_0$ according to Eq~\ref{alpha_a}, as well as with its associated temperatures ratio $T/T_D$. This is useful because allow finding a relation between $\alpha$ and the redshift. The best thermometer to measure the Universe temperature, i.e., how the universe has cooled due to the to expansion is through the cosmic microwave background radiation (CMB). The CMB data, at least up to the redshift of z$\sim$ 3, is consistent with a linear relation, between the temperature and redshift 
\begin{equation}
\frac{T}{T_0}=(1+z),
\end{equation}
where $T_0$ is the temperature at $z=0$, that is, the current temperature, $T_0=2.73\;K$. The last equation can be expressed as
\begin{equation}
\frac{T}{T_D}=\frac{T_0}{T_D} (1+z)=0.43 (1+z),
\end{equation}
at $z=0$ $T_0/T_D=0.43$ (see left panel in Fig.~\ref{alpha_redshift}).Then, $T_D=T_0/0.43=6.35\;K$. The $T_D$ is the only free parameter of DGT, the Debye temperature for the universe is a little more than twice as much of  $T_0$. Fig-\ref{alpha_redshift} (right panel) show the dependence of $\alpha$ as a function of redshift.

\section{Rotation curves}
\label{rotationa}

As above indicated, the current Universe temperature is 2.73 K, this is the temperature of the cosmic microwave background radiation, which permeates the entire Universe, this value is lower than the Debye temperature of Universe, $T_D=6.35$ K, this means that the ratio $T/T_D$ is less than one. In most cases, this condition is expressed in term of the acceleration scale parameter as $a/a_0=6/\pi^2 (T/T_D)$. Thus systems with an acceleration  $a<a_0$ are very well described by MOND theory.  


For nearby galaxies and at large radii, the curves of rotation are ``flat''. This pattern of the rotation completely
escapes the predictions of theories such as the Newtonian, and the General Relativity Theory, which provide for
decreasing rotation curves. To solve this discrepancy is necessary to admit a hidden mass in galaxies. Thus, the main
indirect evidence for dark matter comes from rotation curves of galaxies.


However, for distant galaxies, those with high redshift as the observed by VLT telescopes
with an average redshift of 1.52, means that the Universe temperature was $T=6.88$  K,  with a ratio $T/T_D=1.08$, under this condition the system motion is no longer in the MONDian regime (see Fig.~\ref{debye_function}), in this case, a
declining rotation curve is observed \citep{lang17}. 


In this paper, we present a straightforward analysis on the predictions of DGT
for the rotation curves of galaxies at large radius and in a wide
range of redshifts. 
Equation~\ref{eq:mainDGT} allow to obtain the asymptotic rotation speed as function of the galaxy radius and taking into account the relation for the centripetal acceleration as $a=\mathrm{v}^2/R$, the rotation speed can be expressed
\begin{equation}
\mathrm{v}=(GM a_0^{\alpha})^{1/(2\alpha +2)}R^{(\alpha-1)/(2\alpha+2)}.
\label{eq:speed}
\end{equation}

The deduction of this equation was presented for the first time in our
previous article, in June 2017 [Navia, 2017]. It has two asymptotic limits


$$\mathrm{v}=\left\{\begin{array}{rc}
\sqrt{\frac{GM}{R} },&\mbox{if}\quad \alpha =0 \;\;(T/T_D >> 1),\\
(GMa_0)^{1/4} , &\mbox{if}\quad \alpha=1 \;\;(T/T_D<<1).
\end{array}\right.
$$
The first asymptotic limit happens at high temperatures, and DGT coincides with the deep-Newtonian regime, and this means Keplerian rotation curves.
According to the right panel of Fig~\ref{alpha_redshift}, $\alpha=0$ means high redshifts, i.e., distant galaxies, $z\geq 4$.
The second asymptotic limit corresponds to low temperatures, and DGT  coincides with the deep-MONDian regime, and we have the asymptotically flat rotation curves. According to the right panel of
Fig.~\ref{alpha_redshift}, $\alpha=1$ means very low redshifts,  nearby galaxies, i.e., those with redshift close to zero.  

\begin{figure}
\vspace*{-0.0cm}
\hspace*{0.0cm}
\centering
\includegraphics[width=13.0cm]{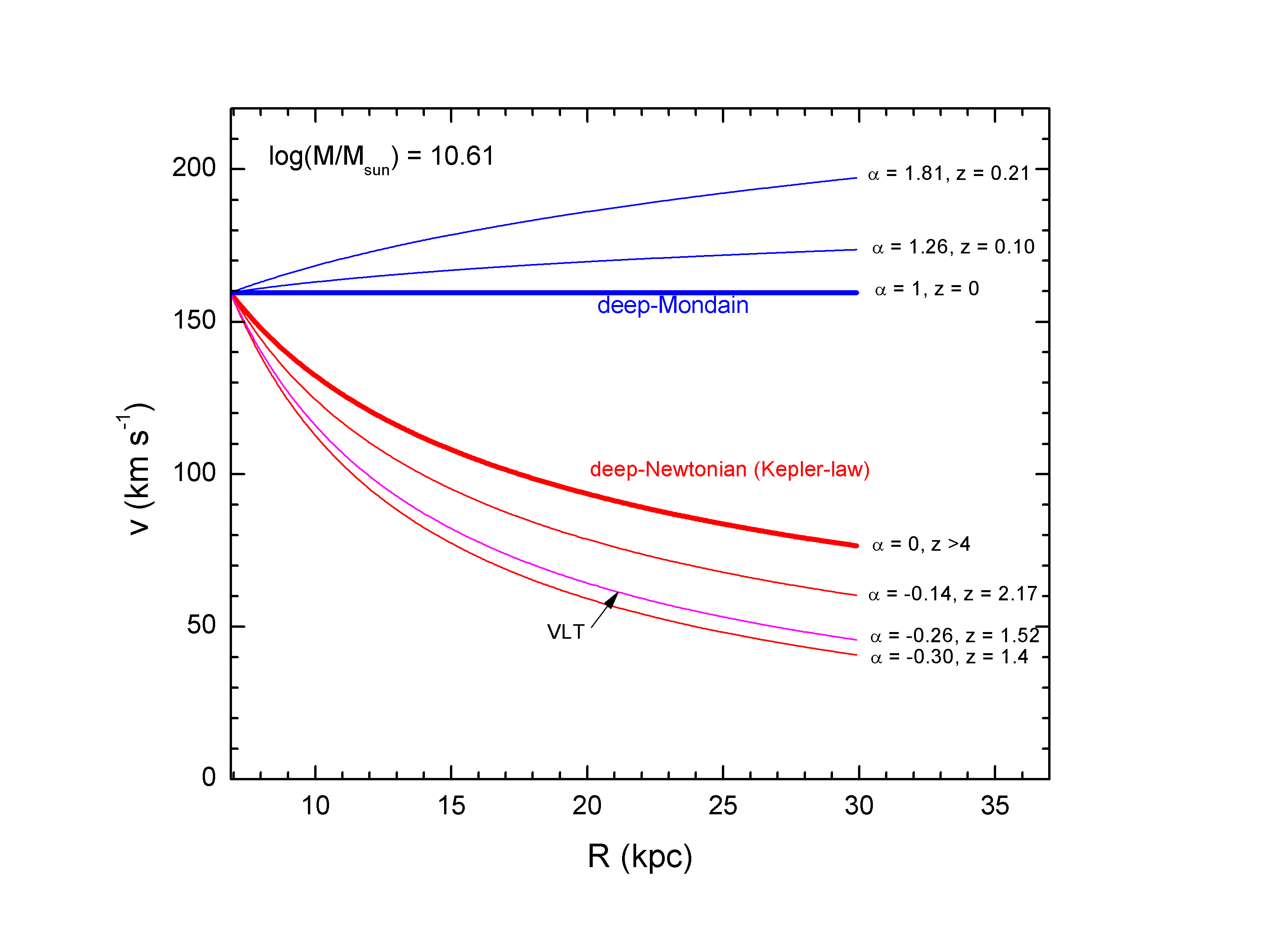}
\vspace*{-0.0cm}
\caption{ DGT predictions for the galaxies rotation curves at large radii, to several redshifts.
In all cases is assumed a mass of $M_G =4.07 \times 10^{10} × M_{Sun} $,
this value coincides with the average mass value of the VLT galaxies.
The red lines represent galaxies with redshift above 0.77, with declining rotation curves (Newtonian regime), in this case, the red bold line indicating
the deep-Newtonian regime (Kepler-law).
The blue lines represents those with redshift below 0.77, with rising rotation curves (MONDian regime). The blue bold line indicate the deep-MONDian regime and happens at $z\sim 0$. 
}
\vspace*{0.5cm}
\label{rotation1}
\end{figure}

\begin{figure}
\vspace*{-0.0cm}
\hspace*{0.0cm}
\centering
\includegraphics[width=13.0cm]{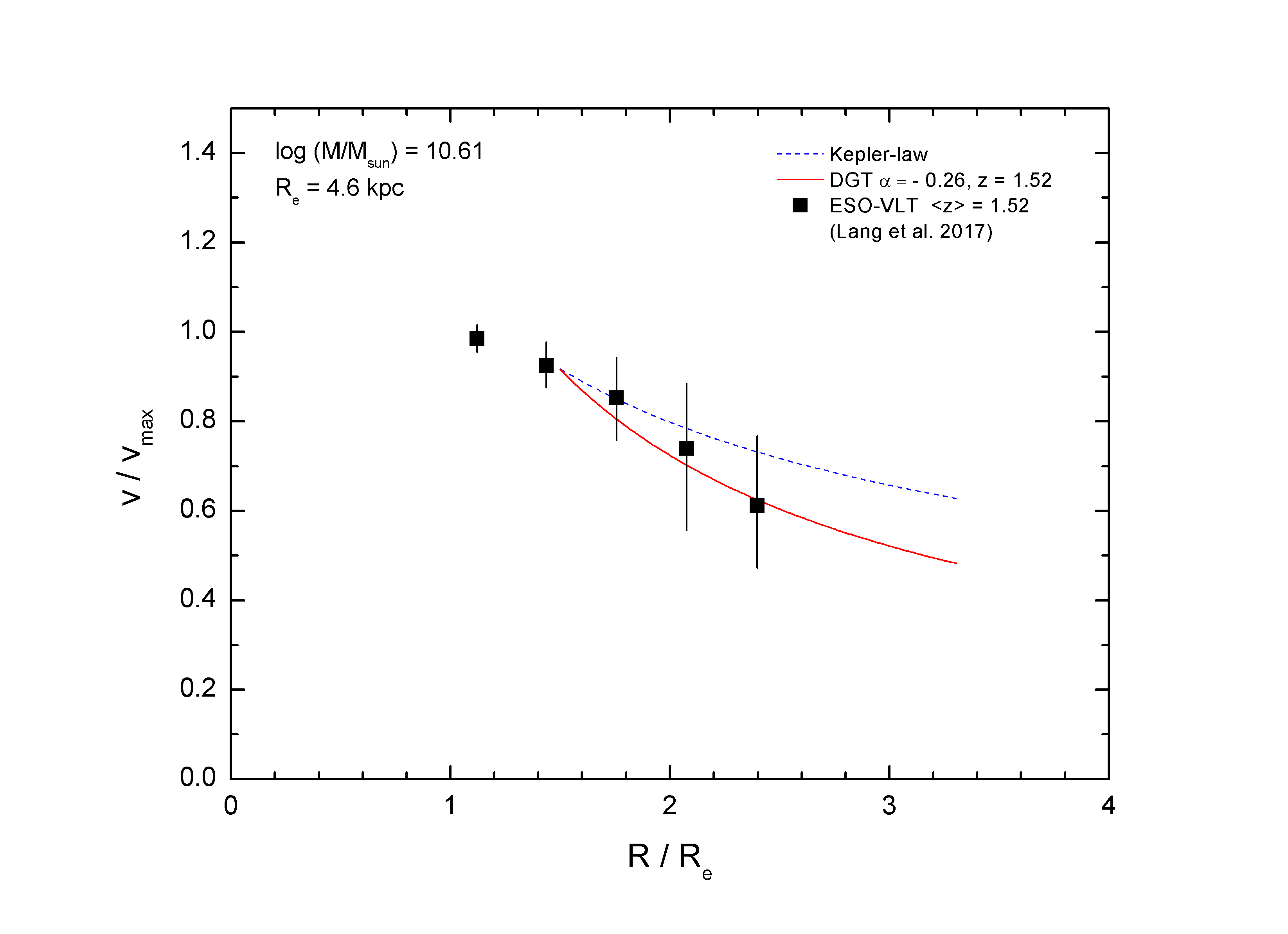}
\vspace*{-0.0cm}
\caption{Comparison among normalized stacked rotation curve at large radii. Black squares represent VLT observations \citep{lang17}, shown together with the expected Kepler-law 
(black dot curve), and DGT prediction to $\alpha=-0.26$ and correspond to redshift $z=1.56$ (red solid curve).
}
\vspace*{0.5cm}
\label{rotation2}
\end{figure}

In DGT, the shape of the rotation curves of the galaxies is determined by the value of the index of R
\begin{equation}
\beta=\frac{\alpha-1}{2\alpha+2},
\end{equation}
that diverges to $\alpha=-1$, this happen at redshift 0.77 (see right panel of Fig.~\ref{alpha_redshift}). Thus the value $z \sim 0.77$ is the limit between two physical regimes. Galaxies with redshift above 0.77 are characterized by a negative $\alpha$ value, in this case, the index $\beta$ is always negative, consequently, the galaxies have declining rotation curves. While galaxies with a redshift below 0.77 are characterized for a positive $\alpha$ value, in this case, $\alpha$ is always higher than one, then the index $\beta$ is always positive, consequently, the rotation curves are not in decline, they are above the flat rotation curve. Fig.~\ref{rotation1} summarize
the situation.

It is intriguing to note that the critical value $ z \sim 0.77$, is close to the redshift, from which the expansion of the Universe began to accelerate, this according to observations of type 1a supernovae \citep{ries98}.

Fig.~\ref{rotation2} shows a comparison between  DGT prediction and the VLT observations \citep{lang17} to the rotation curve of a galaxy at large radii. The DGT prediction is for a galaxy at a redshift that coincides with the VLT average value ($z=1.52$) and to the average mass value of the VLT galaxies.

 \section{DGT predictions for the Radial Acceleration Relation of galaxies}
 \label{radiala}
 
The radial acceleration relation (RAR) is a relation between the observed acceleration $g_{obs}$ and the acceleration
generated by the baryonic mass $g_{bar}$ at any given radius within galaxies, always following the same pattern $g_{obs} \sim \sqrt{g_{bar}}$ to acceleratios below $10^{-10}$ m$s^{-1}$.
The RAR does not care about the specific
properties and morphologies of the galaxies, the relation exists in nearby high-mass elliptical and low-mass spheroidal galaxies. The origin of this scaling-law relation still is unclear. It is claimed \citep{lell17} that RAR is tantamount to a new natural law. The
observed acceleration tightly correlates with the gravitational acceleration from the visible mass only, and it is difficult
to understand in terms of dark matter. However, also is claimed that RAR arises from interactions between baryons
and dark matter \citep{chas18}, or from stellar feedback processes on galaxies \citep{kell17}.
We show that according to DGT, the RAR scaling-law observed in nearby galaxies is broken when are considered
distant galaxies. DGT allows obtaining the RAR in a wide range of redshifts from $z\sim 0$ to $z \sim 4$ because in this range there is a well known linear relation, between the temperature and redshift.


 The observed acceleration in galaxies is obtained from
measurements of the circular velocity at large radii as
\begin{equation}
g_{obs} \sim \frac{V^2}{R},
\end{equation}
while the gravitational acceleration due to the baryonic matter is
\begin{equation}
g_{bar} \sim \frac{GM_{bar}}{R^2},
\end{equation}
Taking into account the Eq.~\ref{eq:mainDGT} we can writing it as
\begin{equation}
g_{obs} \left(\frac{g_{obs}}{a_0}\right)^{\alpha} \sim g_{bar}.
\end{equation}
it can be rewritten  in a more compact form as
\begin{equation}
g_{obs} \sim \left(a_0^{\alpha}g_{bar}\right)^{1/(1+\alpha)}.
\label{DGT_predi}
\end{equation}
that has two limits

\[ g_{obs} \sim
  \begin{cases}
    g_{bar}      & \quad \text{if } \alpha = 0 \text{ (deep-Newtonian)}\\
    \sqrt{a_0 g_{bar}}  & \quad \text{if } \alpha = 1 \text{ (deep-MONDian)}.\\
  \end{cases}
\] 

The DGT prediction, expressed in Eq~\ref{DGT_predi} can be written as
\begin{equation}
\log (g_{obs}) \sim  \frac{1}{1+\alpha} \log (g_{bar}) +  \frac{\alpha}{1+\alpha} \log(a_0),
\end{equation}

In DGT, the shape of the RAR is determined by the value of the slope
\begin{equation}
\delta=\frac{1}{1+\alpha},
\end{equation}

that diverges to $\alpha=-1$, this happens at redshift 0.77. Thus
as already has been observed in the case of the rotation curves, the value $z \sim 0.77$ is the limit between two physical regimes.

Galaxies with redshift below 0.77
have a $\alpha$ value always higher than one, this means
a slope $\delta$ lower than 0.5 and the normalization factor
($\alpha/(\alpha +1)\times \log a_0$) put they above 
the RAR scaling-law line, which happens to $\delta = 0.5$ ($\alpha =1$,
deep-MONDian regime, z $\sim$ 0).
In this case, $g_{obs}$ is systematically higher than $g_{bar}$.

\begin{figure}
\vspace*{-0.0cm}
\hspace*{0.0cm}
\centering
\includegraphics[width=12.0cm]{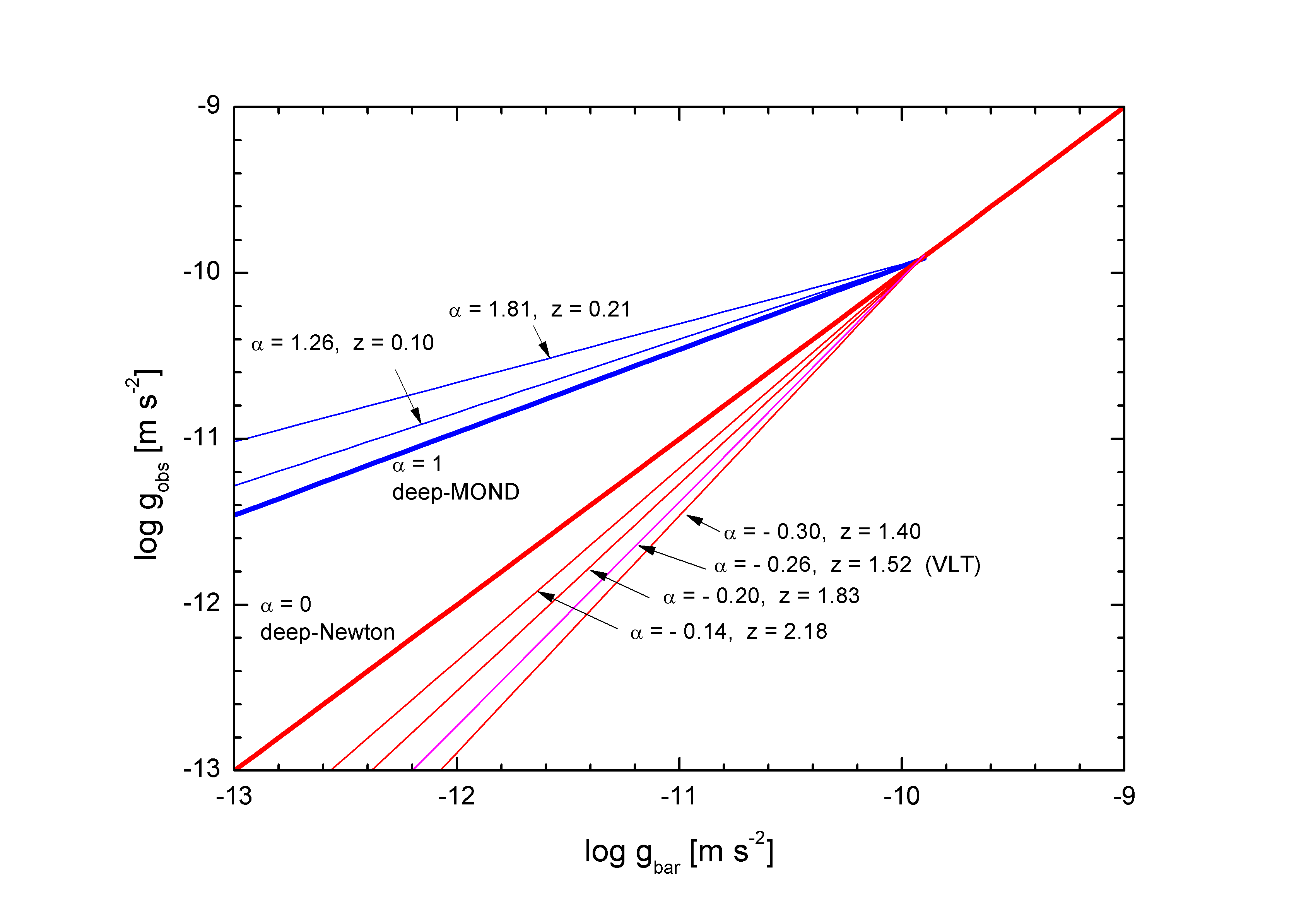}
\vspace*{-0.0cm}
\caption{DGT predictions to the centripetal acceleration observed in rotation
curves of galaxies at large radii, $g_{obs} = V^2/R$, as a function of the
acceleration generated by the baryon mass, $g_{bar} = GM_{bar}/R^2$.
The red lines represent galaxies with redshift above 0.77, with declining rotation curves (Newtonian regime), in this case, the red bold line indicating
the deep-Newtonian regime (Kepler-law) $g_{obs}=g_{bar}$.
The blue lines represent those with redshift below 0.77, with rising rotation curves (MONDian regime). The blue bold line indicates the deep-MONDian regime and happens at $z\sim 0$.
}
\label{radial}
\end{figure}

While galaxies with a redshift above 0.77 have a $\alpha$ value always negative 
(but, higher than -1),
this means that are characterized by a slope $\delta$ higher than one and  the normalization factor
($\alpha/(\alpha +1)\times \log a_0$) put they below 
the deep-Newtonian line, which happens to $\delta = 1$ ($\alpha=0$).  
Thus in this case, $g_{obs}$ is systematically lower than $g_{bar}$. 
Fig.~\ref{radial} summarize the situation.

Recently has been claimed through LCDM Monte Carlo simulations, that the radial acceleration relation (to accelerations below $10^{-10}$ m s$^{-2}$) is a consequence of the dissipative collapse of baryons, rather than being evidence for exotic
dark-sector physics or new dynamical laws \citep{kell17}. This mechanism predicts that of $g_{obs}$ increase
with the redshift, for instance, $g_{obs}$ at z$=$2 must be higher than at z$\sim$ 0 (nearby galaxies). This evolution must be a consequence of the huge impact that stellar feedback has on galaxies. However, according to the ESO VLT observations, galaxies at
redshift z$=$2 have declining rotation curves, below the Keplerian-law curve, thus the rotation velocity decreases faster as
the radius increase, this means that the $g_{obs}$ at z$=$2 must be smaller than at z$=$0. 
Thus, the predictions from LCDM simulations (including feedback processes) to the RAR at high redshifts, are in conflict with the observations.

\begin{figure}
\vspace*{-0.0cm}
\hspace*{0.0cm}
\centering
\includegraphics[width=12.0cm]{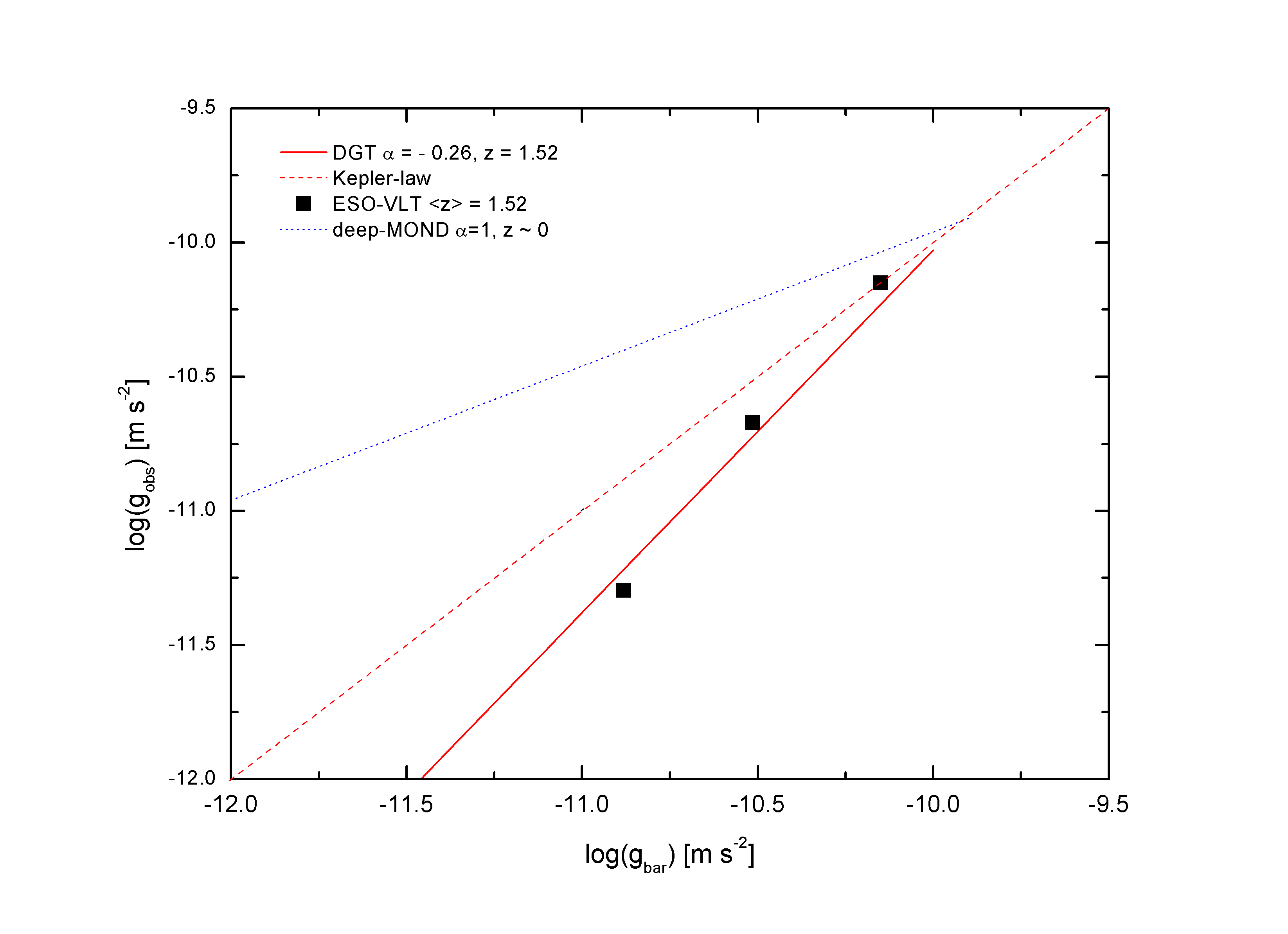}
\vspace*{-0.0cm}
\caption{Comparison among Radial Acceleration Relation at large radii. Black squares represent VLT observations \citep{lang17}, they were obtained from the three points in Fig.~\ref{rotation2}.
Shown together with DGT prediction to $\alpha=-0.26$ and correspond to redshift $z=1.56$ (red solid line), the expected according to Kepler-law 
(red dashed line), and the expected according to deep-MOND  (blue dot line).
}
\label{radial2}
\end{figure}

Fig.~\ref{radial2} shows a comparison between the expected RAR according to DGT for galaxies at redshift 1.52 and the RAR obtained from 
VLT observations of the rotation of galaxies at an average redshift of 1.52. (from the three points in Fig.~\ref{rotation2}).

The above results contrast with those obtained with the
introduction of the so-called, stellar feedback processes in LCDM galaxy-formation simulations.
These calculations allowed reproducing the RAR scaling-law in nearby galaxies \citep{kell17}, but it predicted that is broken, for distant galaxies, for instance, at z$\sim$2, $g_{obs}$ is higher than $g_{bar}$, and $g_{obs}$ increases as the redshift increase.

 This prediction of LCDM (plus feedback processes) are in conflict with the RAR obtained from VLT observations and DGT predictions, at high redshifts (z $\sim$ 2) the RAR scaling-law is broken, but in the opposite direction,   $g_{obs}$ is lower than $g_{bar}$. This result also is a clear challenge to cold dark matter cosmology.

\section{Conclusions}
\label{conclusions}

In this paper, we have followed the concept, initially proposed by Sakharov of ``induced gravity'' \citep{viss02}, and means that gravity is not ``fundamental'' in the sense of particle physics. Instead of that, it emerges from quantum field theory in the same sense that hydrodynamics or continuum elasticity theory emerges from molecular physics.

 In this sense in DGT gravity is induced by the entropy variation of a system constituted by oscillating quasi-particles (information bits) on a closed holographic screen and that stores the information of matter enclosed within it.
 Thus, DGT is not a modified gravity theory, because it does not modify anything. DGT is built from first principles. In DGT, the Entropic Gravity Theory plays the role of Dulong-Petit law, for the specific heat of solids at high temperatures. In this framework Debye develops a theory that explains the specific heat of solids at low temperatures. Analogously, we introduced the Debye scheme in the entropic gravity theory to explain gravity at low temperatures.
DGT has two asymptotic boundaries, at low temperatures it coincides with MOND theory and at
high temperatures DGT coincides with Newtonian gravity, also in this limit, DGT is compatible with GRT.

As gravity in DGT depends on temperature, it is possible to incorporate the redshift of the system in analyze, from z=0 to z $\sim$ 4, because in this range, is well known the linear relation between the temperature and redshift, obtained from observations of the cosmic microwave background radiation.

The main result of the DGT indicates that the temperature of the thermal bath in which the galaxies move regulates its dynamics. DGT finds for the Debye temperature of the Universe the value $ T_D = 6.35 $ K. Galaxies in a thermal bath with a temperature less than $ T_D $ ($z<0.77$) and at large radii have increasing rotation curves, above  ``flat'' rotation curve predicted by deep-MONDian regime, this means that $g_{obs}$ is systematically higher than $g_{bar}$. In this group are the nearby galaxies ($z \sim 0$), in this case, is observed the RAR as a scaling-law with $g_{obs}\sim \sqrt{g_bar}$. 

However, if the thermal bath has a temperature T greater than $ T_D $ ($z>0.77$) and at large radii, the galaxies have falling rotation curves, and the fall is faster than Kepler-law rotation, this means that $g_{obs}$ is systematically lower than $g_{bar}$.
Thus, in DGT the scaling-law observed in nearby galaxies is broken for distant galaxies.
These DGT predictions, the rotation curves, and the RAR are in line with observations from  VLT telescope for an average redshift of 1.55.

We believe to be possible to improve DGT, because even Debye theory of solid
on which it is based DGT has some limitations and divergences, such as the prediction for the thermal conductivity that tends to infinite. One way to avoid these inconsistencies is to include higher order terms in the potential energy, i.e., to include
anharmonic terms in the phonon vibrations. This refinements also can be made on DGT to obtain a better description of astrophysical observations.

\acknowledgments

This work is supported by the Conselho Nacional de Desenvolvimento Cient\'{i}fico e Tecnol\'{o}lgico (CNPq) Brazil. Grant 152050/2016




\end{document}